# Improvement of Critical Current Density and Upper Critical Field in MgB$_2$ using Carbohydrate

J. H. Kim, M. S. A. Hossain, X. Xu, W. K. Yeoh, D. Q. Shi, S. X. Dou, S. Ryu, M. Rindfleisch, and M. Tomsic

*Abstract*— We evaluated the doping effects of carbohydrate (malic acid, C$_4$H$_6$O$_5$), from 0wt% to 30wt% of total MgB$_2$, on the phase, lattice parameters, critical temperature ($T_c$), resistivity ($\rho$), and upper critical field ($H_{c2}$) of MgB$_2$ superconductor. The lattice parameters calculated show a large decrease in the *a*-axis for MgB$_2$ + C$_4$H$_6$O$_5$ samples, but no change in the *c*-axis. This is an indication of the carbon (C) substitution into B coming from C$_4$H$_6$O$_5$, resulting in enhancement of $\rho$, $J_c$, and $H_{c2}$. Specifically, the $J_c$ value of 2.5 × 10$^4$ Acm$^{-2}$ at 5 K and 8 T for the MgB$_2$ + 30wt% C$_4$H$_6$O$_5$ sample is higher than that of the un-doped MgB$_2$ by a factor of 21. In addition, $\rho$ value for all the MgB$_2$ + C$_4$H$_6$O$_5$ samples ranged from 80 μΩ·cm to 90 μΩ·cm at 40 K, which is higher than for un-doped MgB$_2$. The increased $\rho$ indicates increased impurity scattering due to C, resulting in enhanced $H_{c2}$.

## I. Introduction

THE discovery of the superconductivity of MgB$_2$ at 39 K could offer the promise of important large-scale applications at around 20 K [1]. A significant enhancement in $J_c$ of MgB$_2$ has been achieved through chemical doping with carbon (C) containing compounds, such as SiC, C, B$_4$C, and CNT [2-5]. In particular, C can enter the MgB$_2$ structure by the substitution for B to increase impurity scattering into the two-band MgB$_2$ and hence significantly increase $J_c$ and $H_{c2}$. However, the doping effect for MgB$_2$ superconductor has been limited by the agglomeration of nano-sized dopants and poor reactivity between boron (B) and C. Un-reacted C remain in the MgB$_2$ matrix and react with Mg or B powders, resulting in the formation of impurity phases. To overcome these properties, it is further necessary to study the homogeneity of mixing as well as the activity of reaction between B and C materials. The various methods have been studied by many groups. Ribeiro *et al.* reported that B$_4$C was considered to be one of the more reactive C containing compounds because free C was liberated from B$_4$C [6]. However, using B$_4$C compound was needed higher sintering temperature up to 1200°C as well as longer sintering time. Dou *et al.* studied the effect of SiC doping on the MgB$_2$ superconductor [7]. Even though SiC material was much more effective at enhancing the $J_c$ at a low sintering temperature (~650°C), there were agglomerations of nano-particles in the MgB$_2$ matrix.

In this study, therefore, we used malic acid (C$_4$H$_6$O$_5$) as a representative carbohydrate dopant. Because of the high reactivity of freshly formed C, the C substitution for B can take place at the same temperature as the formation temperature of MgB$_2$. Carbohydrates can be dissolved in a solvent (~150°C) so that the solution can form a slurry with B powder. After evaporating the solvent the carbohydrate forms a coating on the B powder surfaces, giving a highly uniform mixture. For these reasons, the doping effects of carbohydrate material on the MgB$_2$ superconductor were evaluated. We fabricated MgB$_2$ superconductor with C$_4$H$_6$O$_5$ included. The phase, lattice parameters, critical temperature ($T_c$), resistivity ($\rho$), $J_c$, and $H_{c2}$ are presented in comparison with the un-doped reference MgB$_2$.

## II. Experimental Procedure

MgB$_2$ pellets were prepared by an *in-situ* reaction process with addition of malic acid, C$_4$H$_6$O$_5$. The amount of C$_4$H$_6$O$_5$ (99%), from 0wt% to 30wt% of total MgB$_2$, was mixed with an appropriate amount of B (99%) in toluene (C$_7$H$_8$, 99.5%). This slurry was dried in vacuum so that the B powder was coated by the C coming from C$_4$H$_6$O$_5$. Since decomposition temperature of C$_4$H$_6$O$_5$ was at around 150°C. This uniform composite was then mixed with an appropriate amount of Mg (99%) powder. These mixed powders were ground, pressed and then sintered at 900°C for 30 min under high purity argon gas. The heating rate was 5°C/min. All samples were characterized by X-ray diffraction (XRD). The crystal structure was refined with the aid of the program FullProf. $T_c$ was defined as the onset temperature at which diamagnetic properties were observed. The magnetization was measured at 5 and 20 K using a Physical Property Measurement System (PPMS, Quantum Design) in a time-varying magnetic field with sweep rate 50 Oe/s and amplitude 8.5 T. Since there is a large sample size effect on the magnetic $J_c$ for MgB$_2$ all the samples for measurement were made to the same size (1 x 2.2 x 3.3 mm$^3$) for comparison [5]. The magnetic $J_c$ was derived from the width of the magnetization loop using Bean's model. In addition, $H_{c2}$ was

Manuscript received August 29, 2006. The authors gratefully acknowledge financial support from the Australia Research Council, Hyper Tech Research Inc, and CMS Alphatech International Ltd.

J. H. Kim, M. S. A. Hossain, X. Xu, W. K. Yeoh, D. Q. Shi, and S. X. Dou are with Institute for Superconducting and Electronic Materials, University of Wollongong, Wollongong, NSW 2522, Australia (e-mail: jhk@uow.edu.au).

S. Ryu is with Center for Scientific Instruments, Andong National University, Andong-si, Gyeongsangbuk-do 760-749, Republic of Korea.

M. Rindfleisch and M. Tomsic are with Hyper Tech Research, Inc., 1275 Kinnear Road, Columbus, OH 43212, USA.

defined as $H_{c2}=0.9R(T_c)$ from the resistance ($R$) versus temperature ($T$) curve.

### III. RESULT AND DISCUSSIONS

Figure 1 shows the XRD result for un-doped MgB$_2$ and MgB$_2$ + C$_4$H$_6$O$_5$ samples. The XRD measurements were performed on the ground MgB$_2$ cores. It was observed that the four different kinds of samples seemed to be well developed MgB$_2$ with a small amount of MgO. XRD patterns of the MgB$_2$ + C$_4$H$_6$O$_5$ sample are almost independent of the sintering temperature. For the MgB$_2$ + 30wt% C$_4$H$_6$O$_5$ sample, however, the relative amount of MgO was slightly larger than for the lightly doped samples. From the XRD results, the full width at half maximum (FWHM) of the (110) and (002) peaks for un-doped and MgB$_2$ + C$_4$H$_6$O$_5$ samples were also evaluated. For example, the FWHMs of the (110) peaks for the un-doped, 10wt%, 20wt%, and 30wt% C$_4$H$_6$O$_5$ samples were calculated to be 0.445°, 0.500°, 0.550°, and 0.535°, respectively, while the corresponding values for the (002) peak were 0.400°, 0.445°, 0.470°, and 0.395°. That is, as the doping level increased to 20wt%, FWHM value linearly increased for both peaks. This significant broadening for the MgB$_2$ + C$_4$H$_6$O$_5$ samples can be explained by lattice distortion due to C substitution from C$_4$H$_6$O$_5$, indicating the degradation of crystallinity of the MgB$_2$ [8]. As the doping level further increased to 30wt%, however, the FWHM value slightly decreased. This observation is related to the improvment of crystallinity.

For further analysis, we calculated the lattice parameters of the $a$- and $c$-axes from the (110) and (002) peaks, respectively, as shown in Figure 2. It was observed that the $a$-axis parameter changed from 3.0835(5) Å to 3.0731(9) Å as the doping level increased to 30wt%. This is attributed to the actual C substitution for B sites. The shrinkage of the $a$-axis occurred due to the difference in atomic size between B and C ions after substitution. On the other hand, the $c$-axis did not change, suggesting that C doping does not affect the $c$-axis. The $c$-axis parameter was in the range of 3.5214(7) Å to 3.5268(3) Å. As a result, C coming from C$_4$H$_6$O$_5$ can be used as a C source, just like SiC, C, B$_4$C and CNT. It is to be noted that the $a$-axis values are almost saturated for the 10wt%, 20wt%, and 30wt% C$_4$H$_6$O$_5$ samples. From the Avdeev et al. results, we also estimated the actual amount of C substitution in our MgB$_2$ + C$_4$H$_6$O$_5$ samples [9]. The value was calculated to be x ~ 0.038 to 0.046 in the composition of MgB$_{2-x}$C$_x$. It is evident that the amount of C substitution for B is much less than the nominal composition. However, even though the actual C levels for C$_4$H$_6$O$_5$ were 1.9at% to 2.3at% of B, the amount of actual C substitution was much higher for other C doped MgB$_2$ sample. This is attributed to the high reactivity of the freshly formed C.

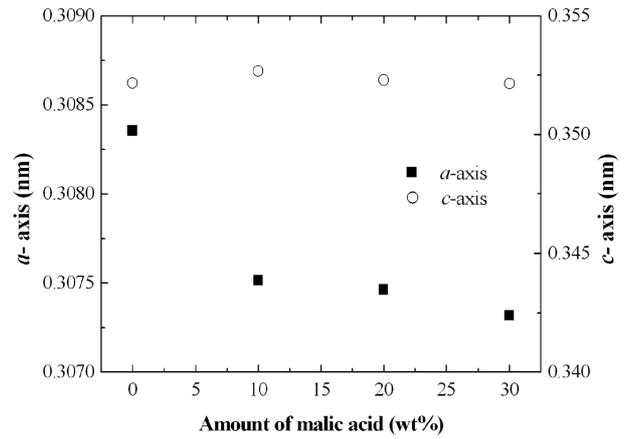

Fig. 2. Lattice parameters $a$- and $c$-axis as a function of the C$_4$H$_6$O$_5$ contents.

Figure 3 shows the $T_c$ of all samples sintered at 900°C for 30 min. The calculated values were 37.6 K, 35.8 K, 35.7 K, and 35.8 K, respectively. Specifically, the $T_c$ behavior of the MgB$_2$ + 10wt% C$_4$H$_6$O$_5$ sample had a two-step behavior near the onset temperature. This is probably related to inhomogeneity of the mixed powder between B and C. In this experiment, using toluene solution is believed to be much more effective for preparing homogeneous powder, since carbohydrate is easy to dissolve in toluene. After decomposition of the carbohydrate, C remaining C can from a coating on the surface of the B powder, resulting in homogeneous mixing. For future work, we need further systematical study on the use of solution methods. In addition, we observed that $T_c$ broadening occurred for the $MgB_2$ + $C_4H_6O_5$ samples. This could be associated with degradation of crystallinity due to C substitution. This result is consistent with the results from the FWHM and lattice parameter.

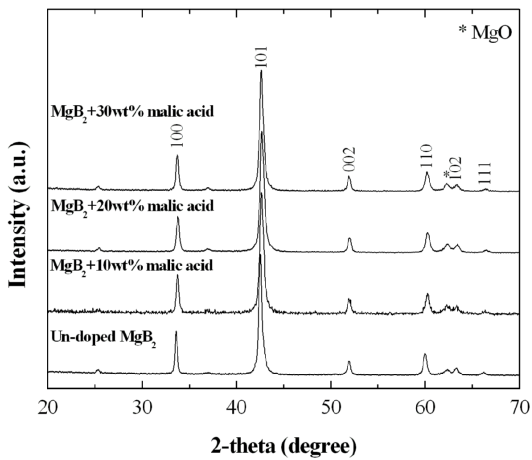

Fig. 1. The x-ray diffraction patters for un-doped MgB$_2$ and MgB$_2$ + C$_4$H$_6$O$_5$ samples with different compositions.

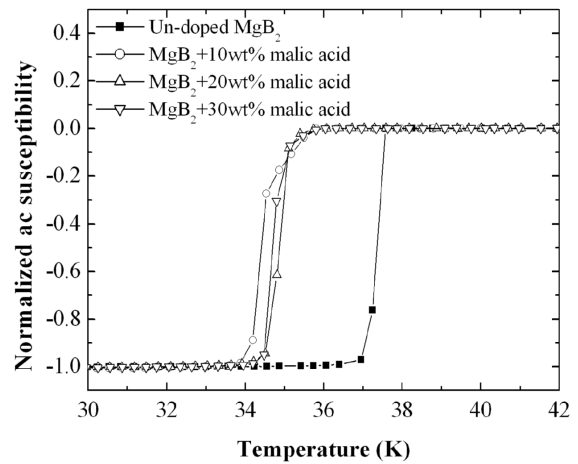

Fig. 3. Critical temperature as a function of the C$_4$H$_6$O$_5$ contents.

Figure 4 shows the magnetic field dependence of $J_c$ in all samples at 20 K and 5 K. It should be noted that $J_c$ values in high field were increased by more than an order of magnitude. For example, the $J_c$ value of $2.5 \times 10^4$ Acm$^{-2}$ at 5 K and 8 T for the $MgB_2$ + 30wt% $C_4H_6O_5$ sample is higher than that of the un-doped $MgB_2$ by a factor of 21. In addition, there was no $J_c$ degradation in self-field for the $MgB_2$ + 30wt% $C_4H_6O_5$ sample. The estimated $J_c$ value was $4.0 \times 10^5$ Acm$^{-2}$ at 20 K and self-field.

The temperature dependence of $H_{c2}$ for all samples is shown in Figure 5. Significantly enhanced $H_{c2}$ for $MgB_2$ + $C_4H_6O_5$ samples was observed, suggesting that C substitution into B sites results in an enhancement in $H_{irr}$ and $H_{c2}$. The $\rho$s for the un-doped and $MgB_2$ + $C_4H_6O_5$ samples are 34 $\mu\Omega$·cm and 80 $\mu\Omega$·cm to 90 $\mu\Omega$·cm at 40 K, respectively. The increased $\rho$ for $MgB_2$ + $C_4H_6O_5$ samples indicates the increased impurity scattering as a result of C substitution into B sites. It is to be noted that RRR values from the formula $\rho(300K)/\rho(40K)$ slightly increased as the doping level increased. These values were 2.13, 1.62, 2.34, and 2.23 for corresponding compositions. These relatively high values of RRR indicate that the samples are of good quality, even the 20wt% and 30wt% doping samples.

As a result, the significant advantages of carbohydrate are: (1) the carbohydrates in the mixture melt at lower temperatures and decompose at temperatures below the formation temperature of $MgB_2$, hence producing highly reactive and fresh C on the atomic scale: and (2) as carbohydrates consist of a large range of materials, this work will have significant implications for further improvement of the performance properties of $MgB_2$, as well as many other C-based compounds and composites.

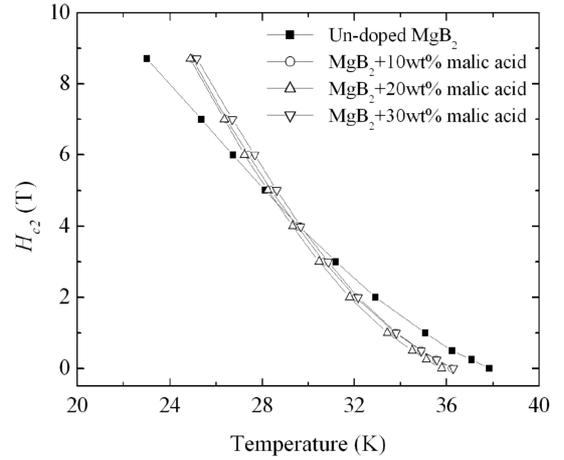

Fig. 5. Upper critical field for un-doped $MgB_2$ and $MgB_2$ + $C_4H_6O_5$ samples.

IV. CONCLUSIONS

We evaluated the doping effects of malic acid, $C_4H_6O_5$, from 0wt% to 30wt% of total $MgB_2$, on the phase, lattice parameters, FWHM, $T_c$, $\rho$, $J_c$, and $H_{c2}$ of $MgB_2$ superconductor. We observed that all samples sintered at 900$^o$C seemed to be well-developed $MgB_2$ with small amounts of MgO. However, the FWHMs related to the (110) and (002) peaks were larger for $MgB_2$ + $C_4H_6O_5$ samples than for un-doped $MgB_2$. This indicates that lattice distortion had occurred due to the C coming from $C_4H_6O_5$. In addition, the $a$-axis parameter changed from 3.0835(5) Å to 3.0731(9) Å as the doping level increased to 30wt%. The actual C substitutions were calculated to be x ~ 0.038 to 0.046 in the composition of $MgB_{2-x}C_x$. Specifically, $MgB_2$ + $C_4H_6O_5$ samples exhibited excellent $J_c$ and $H_{c2}$. This result indicates that strong flux pinning was enhanced by C coming from $C_4H_6O_5$. The carbohydrate is promising C source for $MgB_2$ superconductor with excellent $J_c$ and $H_{c2}$.

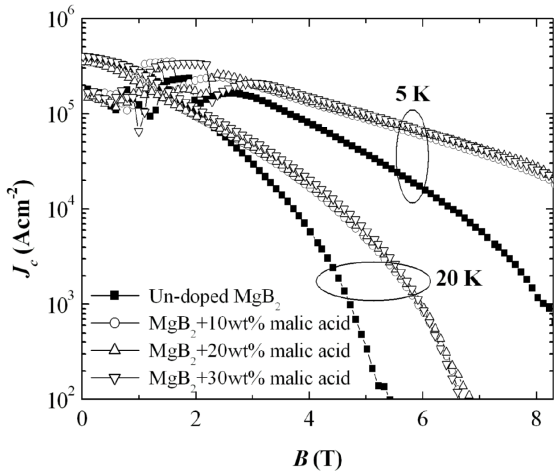

Fig. 4. Magnetic field dependence of critical current density in un-doped $MgB_2$ and $MgB_2$ + $C_4H_6O_5$ samples at 20 K and 5 K.

...